\documentclass[useAMS,usenatbib]{mn2e}
\usepackage{epsfig}
\usepackage{amsmath, amssymb,bm,graphicx}

\title[Asteroid Belts and Life]{On the formation and evolution of
  asteroid belts and their potential significance for life}

\author[R. G. Martin \& M.  Livio]{Rebecca
  G. Martin$^{1,2}$\thanks{E-mail: rebecca.martin@jila.colorado.edu} and
  Mario Livio$^2$\\ $^1$NASA Sagan Fellow, JILA, University of
  Colorado, Boulder, CO 80309, USA\\ $^2$Space Telescope Science
  Institute, 3700 San Martin Drive, Baltimore, MD 21218, USA \\ }

\begin{document}

\date{}

\pagerange{\pageref{firstpage}--\pageref{lastpage}} 
\pubyear{2012}
\maketitle

\label{firstpage}

\begin{abstract}
Suggestions have been made that asteroid belts may be important both
for the existence of life and perhaps even for the evolution of
complex life on a planet. Using numerical models for protoplanetary
discs we calculate the location of the snow line, and we propose that
asteroid belts are most likely to form in its vicinity. We then show
that observations of warm dust in exo-solar systems, thought to be
produced by collisions between asteroids in a belt, indicate that
asteroid belts (when they exist), indeed coincide with the radial
location and the temperature of the snow line. Giant planets form
outside the snow line and prevent planet formation just inside of
their orbit creating an asteroid belt there. However, the migration of
giant planets through the asteroid belt likely disperses the compact
formation. We examine existing observations of giant exo-planets and
find that less than $4\%$ are at radial locations outside of the snow
line.  This definitely may be the consequence of observational
selection effects. However, with this caveat in mind, we point out
that the dearth of giant planets outside the snow line may also
suggest that compact asteroid belts are not common, and more
speculatively that complex life may not expected in most of the
currently observed systems.
\end{abstract}


\section{Introduction}

\vspace{-0.1cm}

Asteroid belts can (in principle at least) affect the habitability of
a planet, the emergence of life, and potentially even the evolution of
complex life in several important ways. Terrestrial planets typically
form in a dry region of a protoplanetary disc \citep{martin12c}.
Water therefore must be later delivered to the planet's surface, most
likely by asteroids \citep[e.g.][]{morbidelli00}, although other
sources, such as comets and the interaction between the magma and
atmosphere have been suggested \citep{lunine06,genda08}. The formation
of large moons may also require an asteroid collision \citep{canup01},
although here again, a different origin for the impactor has been
considered \citep[e.g.][]{belbruno05}. Our Moon, for instance,
stabilises the rotation axis of the Earth and prevents weather
extremes that would have resulted from chaotic motion. This process
may not be universal, since it depends on initial
conditions. According to some hypotheses, life itself may have been
delivered to Earth by an asteroid \citep[e.g.][]{cronin83,
  castillo08,houtkooper11}. Heavy elements, including some that are
essential for life, were also probably delivered to the Earth's crust
through collisions \citep{willbold11}.  During the early times of
formation the Earth was molten and its gravity pulled heavy elements
to its core leaving the crust depleted of elements such as iron, gold
and platinum. On a much more speculative note, the asteroid impact
that led to the extinction of the dinosaurs on Earth \citep{alvarez80}
may have allowed for the emergence of dominant mammals and intelligent
life. The point is that even if only one of these ideas turns out to
be true, it makes it extremely intriguing to examine the formation and
evolution of asteroid belts and consider the possibility for complex
life in other observed solar systems.

The asteroid belt in our solar system is located between the inner
terrestrial planets and the outer giant planets, between Mars and
Jupiter. It contains millions of irregularly shaped bodies made of
rock, ices and metals with a total mass of about 4 per cent of that of
the Earth's moon.  Studying the asteroid belt allows us a glimpse into
the early stages of planet formation.  Observations suggest that at
the time of planetesimal formation the location of the snow line was
within the asteroid belt. The snow line marks the radius outside of
which ice forms. The inner asteroids, closest to Earth, are water
devoid, while outside a radius of about $2.7\,\rm au$ the asteroids
are icy C-class objects \citep[e.g.][]{abe00,morbidelli00}.

The asteroid belt formed from planetesimals within the solar
nebula. In the generally accepted scenario, planetesimals stuck
together to form larger and larger bodies until the protoplanets were
formed. It is thought that the asteroid belt exists because
gravitational perturbations from Jupiter gave the planetesimals inside
of its orbit too much orbital energy to form a planet. The violent
collisions produced fragmentation rather than fusion
\citep[e.g.][]{edgar04}. The picture that we envisage and examine in
the current work, that was first observationally motivated by
\cite{morales11}, is the following: Giant planets likely form by core
accretion outside of the snow line where the solid mass density is
much higher because of water ice condensation
\citep{pollack96}. Consequently, we propose that if an asteroid belt
forms at all, {\it its location is likely to be around that of the
  snow line}.  Simulations of the formation of terrestrial planets
agree with this scenario \citep[e.g.][]{raymond09}. The habitable,
rocky terrestrial planets should form inside the snow line, as was the
case in our own solar system.

Observational evidence suggests that Jupiter formed close to its
current location. The generally accepted theory of the formation of
our solar system infers that Jupiter migrated through planetesimal
interactions only by about $0.2-0.3\,\rm au$
\citep[e.g.][]{morbidelli10}. The asteroid belt is thought to have
originally contained about an Earth mass.  As Jupiter migrated
slightly inwards, its gravitational perturbations caused the majority
of the mass to be ejected, leaving behind only about 0.1 per cent of
the original. Jupiter is still responsible for nudging asteroids and
sending them towards the inner solar system. It is unlikely that
complex life would have have been able to develop on Earth had the
asteroid belt remained as massive as it had originally been, because
there would have been far too many devastating impacts. However, it is
probably also unlikely that complex life on Earth would exist had all
of the asteroids been scattered away.

To test our suggestion for the location of asteroid belts, in
Section~\ref{obs}, we first use detailed protoplanetary disc models to
examine the location of the snow line for different stellar masses. We
then describe and tabulate the observations of warm dust that may
indicate the presence of exo-asteroid belts and show that their
locations coincide with the snow line.  The formation of the asteroid
belt depends on the formation of an exterior giant planet. In
Section~\ref{hab} we therefore consider observations of giant planets
in exo-solar systems and discuss the implications of their locations
for the presence of terrestrial planets in the habitable zone and the
likelihood of a compact asteroid belt. We also examine the question of
how special our solar system is.

\vspace{-0.6cm}

\section{Asteroid Belts and the Snow Line}
\label{obs}

In this Section we calculate numerical models of protoplanetary discs
to determine the snow line evolution. We then examine observations of
warm dust that may indicate the presence of exo-asteroid belts and
show that their radial locations coincide with the theoretical radius
of the snow line.

\subsection{Theoretical Models of the Snow Line}

The snow line marks the distance from the star outside of which ice
forms. It is thought to occur at temperatures of around $170\,\rm K$
\citep[e.g.][]{lecar06}. \cite{marseille11} find that there is an
extended region down to almost $100\,\rm K$, the snow border, in which
both icy and dry planetesimals can coexist. Particles migrating
through the disc accumulate near the snow line over a short radial
extent and grow through collision \citep{kretke07}.  Planetesimal
formation rates increase by an order of magnitude or more moving
across the snow line when the solid surface density increases by a
factor of two \citep{chambers10}.

To model the evolution of the snow line we used a detailed model for a
layered protoplanetary disc of the type first described in
\cite{armitage01} and further developed in \cite{zhu10} and
\cite{martin11}.  We should note that our model is different from
previous models that assumed a fully turbulent disc
\citep[e.g.][]{oka11}, and failed to explain the current location of
the snow line in the solar system \citep[see][for a
  discussion]{martin12c}. In our model, the magneto-rotational
instability (MRI) drives turbulence and angular momentum transport
within the disc if it is sufficiently ionised
\citep{balbus91}. However, protoplanetary discs are thought to contain
dead zones at the mid-plane (a region of zero MRI turbulence), because
they are too cool for thermal ionisation, and cosmic rays or X-rays do
not penetrate the entire disc \citep{gammie96}. We used a critical
magnetic Reynolds number, $Re_{\rm M,crit}=5\times 10^4$, to determine
the precise extent of the dead zone \citep[e.g. as
  in][]{martin12a,martin12b}. In our simulations, where a dead zone is
present, material accumulates and flows only through the active layers
on the surfaces. Once sufficient material collects, the dead zone
becomes self-gravitating and a second type of gravitational turbulence
is driven. The increased heating leads to the MRI being triggered in
the dead zone and FU Orionis type outbursts. In the present work we
are interested in the detailed structure of the disc at late
evolutionary times, after the outbursts have ceased.

\begin{figure}
\centerline{\includegraphics[width=6.5cm]{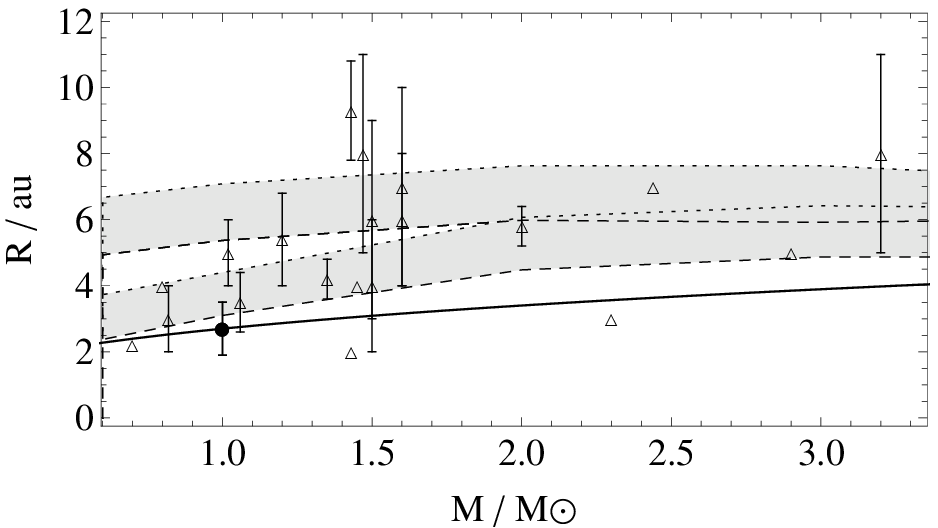}}
\centerline{\includegraphics[width=6.5cm]{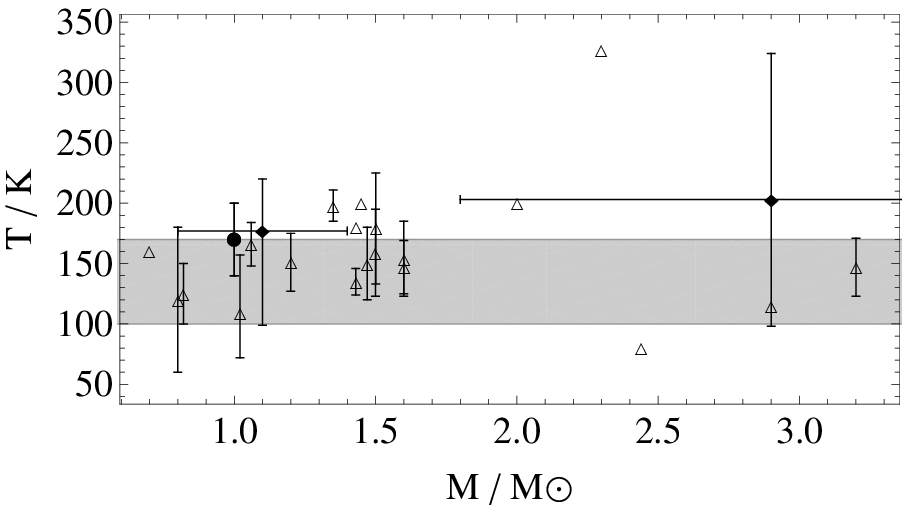}}
\caption{Upper panel: Radius of the observed warm dust. The shaded
  regions show the snow border found from numerical models at times
  $t=10^6\,\rm yr$ (upper) and $t=10^7\,\rm yr$ (lower).  The solid
  line shows the analytic approximation to the snow line given in
  equation~(1).  Lower panel: Temperature of the observed warm
  dust. The shaded region shows the snow border. In both plots, the
  open triangles show the individual systems shown in Table~1. The
  filled diamonds (only in the temperature plot) show two samples from
  \protect\cite{morales11} and the error bars show the range in the
  samples. The filled circles show the location of our solar system's
  snow line and the range shows the extent of our asteroid belt.}
\label{surf}
\end{figure}

We found that the snow line of the disc moves in over time \citep[as
  was also shown in][]{martin12c}.  Here, we use the same disc model
as in \cite{martin12c} but we expand the calculation and consider
central stars of different mass, $M$.  For each stellar mass we
numerically evolve the time-dependent disc model up to a time of
$10^7\,\rm yr$, and in Fig.~\ref{surf} we show the radius, $R$, of the
inner and outer edges of the snow border that correspond to
temperatures of $T=170\,\rm K$ and $100\,\rm K$ at times $t=10^6$ and
$10^7\,\rm yr$.  We find that the location of the snow line is only
weakly dependent on the mass of the star. We can heuristically
understand the general shape of the snow line location as a function
of stellar mass by the following simple argument. Since the generated
power is due to accretion, $T^4 \propto M/R^3$.  Scaling to the radial
location of the snow line in our solar system we have
\begin{equation}
R_{\rm snow} \approx 2.7 \left(\frac{M}{M_\odot}\right)^{1/3} \,\rm au.
\end{equation}
Fig.~1 shows that this analytic expression for the shape of the snow
line is a reasonable approximation for the minimum snow line radius
for the numerical models considered in this work.

\begin{table*}
\centering
\begin{tabular}{lcccccccccccc}
\hline
\hline
Source ID & Name & Spectral Type  & $M$    &  $R_{\rm dust}$ & $T_{\rm dust}$ & Age & Reference  \\
    & &&    ($M_\odot$)    &  ($\rm au$) & (K) & (Myr) \\
\hline
\hline
HD 12039 &                & G3/5V       & 1.02     &  $4-6$   & 109  & 30  & $[1]$ \\ %wyatt07
HD 13246 &                & F8V         & 1.06     &  $3.5\pm 0.9$  & $166\pm 18$ & 30 & $[2]$ \\
HD 15115 &                & F2          & 1.5      &  $4\pm2$ & $179 \pm 46$ & 12 &$[3]$\\
HD 15745 &                & F0          & (1.6)    & $6\pm2$  & $147\pm22$ & 12  &$[3]$ \\
HD 16743 &             & F0/F2III/IV & (3.2)& $8\pm3$  &$ 147 \pm 24$ & $10-50$  &$[3]$\\ 
HD 22049 & $\epsilon$ Eri &  K2V        & 0.82     & $3\pm 1$ & $100-150$ &  850 & $[4]$\\
HD 30447 &                & F3V         & (1.5)    &  $6\pm3$ & $159\pm 36$ & 30  &$[3]$\\
HD 38678 & $\zeta$ Lep    & A2 IV-V(n)  & 2.3      & 3        & 327  & 231 &  $[5]$\\
%**HD 39060 & $\beta$  Pic   &   A6V       & 1.75     &  $<5$    & & 12 & $[5]$ \\
HD 53143 &                &G9V/K1V      & 0.8      & 4        & $120 \pm 60$  & 1&   $[6]$\\
HD 53842 &                & F5V         & 1.20     & $5.4\pm 1.4$  & $151 \pm 24 $ & 30 &  $[2]$ \\
HD 86087 &  HR 3927       &   A0V       & 2.44     & 7        & 80 & 50 & $[6]$ \\
HD 98800 &                &    K4/5V    & (0.7)    & 2.2      & 160 &  10   &$[7]$ \\ %wyatt07
HD 109085 & $\eta$ Corvi    &  F2V        & 1.43     & 2        &180 &1000  &$[6]$\\ %wyatt says no
%HD 109573 &               &  A0V        & 2.18     & $<20$    & &8&**(binary)\\
HD 113766 &               &  F3/F5V     & (1.5/1.4)& 4        & 200   &16  &$[6]$\\
%**HD 125162 & $\lambda$  Boo & A0p      & 1.66     & 24  & $120\pm20$  &310&\cite{chen09} **(peculiar star)\\
%**HD 139664 &             &  F5IV-V     &1.29      & 15 & $100\pm60$ &200&\cite{chen09}\\
HD 152598 &               & F0V         & 1.43     & $9.3 \pm 1.5$  & $135 \pm 11$ & $210 \pm 70$ & $[2]$ \\
HD 169666 &               &  F5         & 1.35     & $4.2 \pm 0.6 $ & $198\pm 13$ & $2100$ & $[2]$ \\
HD 172555 &  HR 7012      & A5 IV-V     & 2.0     & $5.8\pm0.6$  &  200  &  12      & $[6]$ \\
HD 181296 & $\eta$  Tel   & A0Vn        & 2.9      & 5        & 115 &12  &  $[6]$\\
HD 192758&                & F0V         &(1.6)     &  $7\pm3$ & $154\pm31$ & 40  &$[3]$\\
%**HD 216956 & Fomalhaut     &  A4V        &2.1       & $<10$    &&200 & $[9]$\\
HD 218396 &  HR 8799      &  A5V        & 1.5      & $8\pm 3$        & $150\pm 30$ & $30-160$ & $[8],[9]$ \\
\hline
\multicolumn{4}{|l|}
{Samples from \cite{morales11}, median values (range)} \\
\hline
19 Solar-type Stars  &   & G0V (K0V-F5)  & (1.1  ($0.8 - 1.4$)) &   &  177 ($99-220$)  & 270 ($40-900$) & \\
50 A-Type Stars &   & A0V (B8 -A7)  & (2.9  ($1.8 -3.8$)) &   &  203 ($98-324$)  & 100 ($5-1000$)  & \\
\hline
\end{tabular}
\caption{Observations of warm dust belts that may be exo-asteroid belts.% **Excluded from our sample based on
%  thespectral type.  
[1] \protect\cite{hines06};
[2] \protect\cite{moor09};
[3] \protect\cite{moor11};
[4] \protect\cite{backman09};
[5] \protect\cite{moerchen10}; 
[6] \protect\cite{chen06};
[7] \protect\cite{low05}; 
[8] \protect\cite{chen09};
[9] \protect\cite{moromartin10}. The masses in brackets have been derived from the spectral type.
}
%[5] \protect\cite{wahhaj03}
%[9]\protect\cite{kalas05}
\label{table1}
\end{table*}

\subsection{Observations of Exo-Asteroid Belts}

\vspace{-0.1cm}

Dusty discs thought to be analogous to the asteroid belt in our solar
system have been inferred around main-sequence stars from their
thermal infrared emission. Surveys from space-based mid- and
far-infrared observations suggest that around 10 to 30 per cent of
main sequence stars have an infrared excess \citep{lagrange00}. Most
of these show colour temperatures of those expected for cold dust in
our Kuiper belt \citep{meyer07}.  However, a small fraction also show
inner warm dust that may indicate the presence of a (more massive)
asteroid-belt-like structure. We note that because we only see the
dust and not the larger asteroids, there may be other explanations for
its presence. For instance, the dust may be the result of planet
formation \citep{kenyon04} or of stochastic events such as the period
of late heavy bombardment in our solar system
\citep{wyatt07}. Consequently, it is not known how common asteroid
belts truly are \citep{chen09}. For example, our own asteroid belt is
below the current limit of observability in exo-solar systems.

In Table~1 we show all the debris discs we have found in the
literature that have a warm component, that could be interpreted as an
asteroid belt, with a measured temperature and inferred radius.  We
also include the median temperature values for two samples from
\cite{morales11} for which radii have not been determined.  We do not
include dust with very high temperatures ($\gtrsim 400\,\rm K$) as
this is more likely to represent a planetary collision
\citep[e.g][]{rhee08}. Fig.~1 shows that the observed warm dust belts
have temperatures similar to the temperature of the snow line
suggesting that they coincide with the snow line. The inferred radii
also agree well with our numerical models of the snow line. The
observations therefore give strong support for our proposed scenario.

\vspace{-0.6cm}

\section{Giant Planets and Habitability}
\label{hab}

The presence of an asteroid belt in an exo-solar system requires a
giant planet. In this Section we first examine the current
observations of exo-solar giant planets and analyse how special our
own solar system is. We then consider how giant planet migration
affects 
%the formation and evolution of 
asteroid belts, terrestrial planets, and the possibility for life.

\vspace{-0.5cm}

\subsection{Giant Planet Observations}

\begin{figure}
\begin{centering}
\centerline{\includegraphics[width=6.5cm]{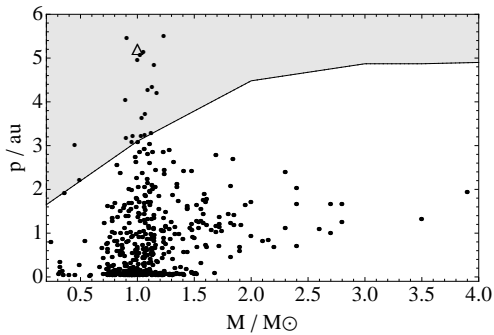}}
\end{centering}
\caption{The distribution of observed giant planet periastron
  separation, $p$, against the mass of the central star, $M$. The open
  triangle shows where Jupiter lies. The shaded region shows the icy
  region outside of the lower limit to the snow line predicted by our
  numerical models.}
\label{data}
\end{figure}

In Fig.~\ref{data} we show the periastron separation against the mass
of the central star for the currently observed giant planets (those
with masses $M>10 M_{\earth}$) in the database exoplanets.org
\citep{wright11}. We shade the region outside of the snow line.  The
boundary represents the lower limit of the distances obtained in the
numerical snow line models (the lowest dashed line in Fig~1).  A large
number of Jupiter-mass planets exist much closer to their host star
than Jupiter is to our Sun.  It is thought that they formed farther
out in the gas disc (outside of the snow line) and migrated inwards
through type II migration \citep[e.g.][]{lin86}. This must have
occurred within a few Myr, while the gas disc was still present
\citep{haisch01}. In order for a planet not to migrate we require it
to form at just the right time (when the gas is depleted) during the
evolution of the protoplanetary disc.  \cite{armitage02} found that
only in about $1-2\%$ of systems does the most massive planet around a
solar type star linger around Jupiter's orbital distance.

During the migration process, a significant fraction of an asteroid
belt (if present) would be accreted on to the star, the planet, or
scattered to large radii \citep[e.g.][]{fogg07}. Hence, we would not
expect a substantial or compact asteroid belt in these systems. The
observed warm dust belts described in Section~\ref{obs}, therefore,
likely remain because there has been little or no giant planet
migration. Giant planets in these systems must have formed towards the
end of the lifetime of the protoplanetary disc.  We do however note
that under special circumstances there could be asteroid belts
elsewhere in the system but statistically speaking the most likely
place that we have identified is around the snow line.

Out of 520 giant planets shown in Fig.~2, only 19 are outside the snow
line. We therefore suggest that less than $4\%$ of these {\it
  observed} systems could possibly harbour a compact asteroid
belt. This is an upper limit because we have taken the lower limit on
the snow line radius (see Section 2) and it is not clear how far out
from the snow line giant planets form. For example, Jupiter is twice
the radial distance of the snow line in our solar system and is
thought to have only migrated a short distance. Planets close to the
snow line, but still outside it, may have undergone more significant
migration than Jupiter which could have disrupted their asteroid belt.
The observational results therefore agree with the theoretical models
of \cite{armitage02} and suggest that our solar system may be rather
special. This argument is based on the detected giant planets, and not
on the (clearly difficult) direct detection of asteroid belts. We
should note though that selection effects almost certainly affect the
statistics, because planets with higher orbital separations are more
difficult to detect \citep[see][for a
  discussion]{cumming08}. Therefore, observations over longer periods
of time will be needed to assess accurately how special our solar
system really is \citep[see also][]{beer04}.

\vspace{-0.4cm}

\subsection{Habitable Terrestrial Planets with a Hot Jupiter?}

Mars-size bodies are formed through planetesimal accretion on a
timescale of $10^5\,\rm yr$ \citep{kokubo98}. These coagulate on a
timescale of around $30-100\,\rm Myr$ to form Earth mass planets
\citep{chambers01}. This is considerably longer than the timescale to
form the giant planets outside the snow line. Thus, the inner giant
planets (``hot Jupiters'') must have migrated through the habitable
zone \citep[that is around $0.95-1.37\,\rm au$ in our solar
  system,][]{kasting93} before terrestrial planet formation there was
complete.  Earth mass planets could still be found (in principle at
least) in orbits inner to the hot Jupiters, however the hot conditions
close to the star are unlikely to produce a habitable planet.

Theoretically, terrestrial planet formation in the habitable zone
after the inward migration of a giant planet is possible
\citep{raymond06,mandell07}. Replenishment of the solid disc
material after the migration is inefficient \citep{armitage03}, but
\cite{fogg07b} found that more than 50 per cent of the solids survive
the migration process and are scattered into larger radius orbits,
where terrestrial planet formation can resume. The solid material of
the scattered disc is diluted, excited and radially mixed. The
resulting planets, that may form within the habitable zone, may be
more water rich than Earth due to inward mixing from outside the snow
line. However, on the basis of the considerations presented in the
Introduction, one may speculate that the lack of a compact asteroid
belt exterior to the planet's orbit makes the probability for complex
life lower.

\vspace{-0.6cm}

\section{Conclusions}

An asteroid belt may be crucial for the emergence and evolution of
life on a planet. Asteroid collisions can deliver water, heavy metals
and possibly even primitive life.  They can also create large moons
which can stabilise planets. While these ideas are clearly
speculative, they are sufficiently intriguing to warrant an
investigation of the formation and evolution of asteroid belts. We
have proposed that asteroid belts, when present, are most likely to
approximately coincide with the location of the snow line in
protoplanetary discs. Using detailed numerical models of
protoplanetary discs we have shown that the radial distance and
temperature of the snow line are indeed entirely consistent with that
of observed warm dust, that could correspond to exo-asteroid belts.

Giant planets are likely to form outside the snow line, where the
solid density increases, thus creating an interior asteroid belt
around the snow line where the planetesimals are too excited to
fuse. However, less than about $4\%$ of the observed giant planets are
found in this region.  Inward migration disrupts the asteroid
belt. Terrestrial planets may still form from scattered planetesimals
in the habitable zone after the migration, but if asteroid belts are
indeed necessary for complex life to evolve, then such evolution is
unlikely in these systems. On the other hand, a small amount of giant
planet migration may be necessary to remove a significant fraction of
the initial belt mass because otherwise there would be too many
devastating impacts on the planet for life to evolve. Consequently,
there appears to be a very narrow ``window of opportunity'' of time
during which the giant planet should form, in order for the correct
amount of migration to take place -- potentially making our solar
system even more special.

We do note, however, that the presently existing data may be
significantly affected by selection effects. Long period planets are
harder to observe, and in the future more giant planets may be
observed outside of the snow line. Based on our scenario, {\it it is
  on these systems (with a giant planet outside of an asteroid belt)
  that we should concentrate our efforts to look for complex life.}
Finally, we note that the asteroid belt could prove to be even more
important to intelligent life on Earth in the future. On one hand,
asteroids could become a source of heavy metals that could be brought
back to Earth should our own sources become depleted
\citep[e.g.][]{lewis97} and in addition, asteroids could also be used
as a launch pad for further exploration. On the other hand, clearly
asteroid impacts present a danger for the continued evolution of
humans on Earth.

\vspace{-0.6cm}

\section*{Acknowledgements}

We thank Christine Chen, Mark Wyatt and Jim Pringle for useful
conversations. RGM thanks the Space Telescope Science Institute for a
Giacconi Fellowship.  RGM's support was provided in part under
contract with the California Institute of Technology (Caltech) funded
by NASA through the Sagan Fellowship Program. This research has made
use of the Exoplanet Orbit Database and the Exoplanet Data Explorer at
exoplanets.org.

%\clearpage

\label{lastpage}
\end{document}